\documentclass[aps,prd,preprint,superscriptaddress,floatfix,nofootinbib]{revtex4-1}

\pdfoutput=1

\usepackage{amsfonts}
\usepackage{epsfig}
\usepackage{bm}
\usepackage{amssymb}
\usepackage{amsmath}
\usepackage{color}
\usepackage{subfigure}

\newcommand{\be}{\begin{equation}}
\newcommand{\ee}{\end{equation}}
\newcommand{\bea}{\begin{eqnarray}}
\newcommand{\eea}{\end{eqnarray}}

\newcommand{\htau}{\hat\tau}

\begin{document}

\title{Analytical attractor and the divergence of the slow-roll expansion in relativistic hydrodynamics}
\date{\today}
\author{Gabriel S.~Denicol}
\affiliation{Instituto de F\'isica, Universidade Federal Fluminense, UFF, Niter\'oi,
24210-346, RJ, Brazil}
\author{Jorge Noronha}
\affiliation{Instituto de F\'isica, Universidade de S\~ao Paulo, Rua do Mat\~ao, 1371,
Butant\~a, 05508-090, S\~ao Paulo, SP, Brazil}

\begin{abstract}
We find the general analytical solution of the viscous relativistic
hydrodynamic equations (in the absence of bulk viscosity and chemical
potential) for a Bjorken expanding fluid with a constant shear viscosity
relaxation time. We analytically determine the hydrodynamic attractor of
this fluid and discuss its properties. We show for the first time that the slow-roll expansion,
a commonly used approach to characterize the attractor, diverges. This is
shown to hold also in a conformal plasma. The gradient expansion is found to
converge in an example where causality and stability are violated.
\end{abstract}

\maketitle



\section{Introduction}

Relativistic hydrodynamics has played a key role in our understanding of the
novel properties of the quark-gluon plasma (QGP) formed in ultrarelativistic
heavy ion collisions (for a review, see \cite{Heinz:2013th}). The basic
picture is that the hot and dense matter formed in these collisions behaves
as a relativistic fluid in which dissipative effects are surprisingly small
in comparison to other fluids in nature \cite{Adams:2012th}. However, recent
experimental observations \cite%
{Abelev:2012ola,Aad:2012gla,Adare:2013piz,Khachatryan:2015waa,Adare:2015ctn,Aad:2015gqa,Khachatryan:2016txc}
have suggested that the strongly interacting matter produced in small
collision systems (such as proton-nucleus and even proton-proton collisions)
also displays the same liquid-like properties found in large nucleus-nucleus
collisions. This finding was accompanied by a number of theoretical studies
on the emergence of hydrodynamic behavior from microscopic models (see, for
instance, \cite%
{Chesler:2009cy,Chesler:2010bi,Heller:2011ju,Casalderrey-Solana:2013aba,Florkowski:2013lza,Denicol:2014xca,Denicol:2014tha,Chesler:2015bba,Heinz:2015gka,Attems:2016tby,Romatschke:2017vte,Florkowski:2017olj}%
), which have contributed to assess the domain of applicability of
relativistic hydrodynamics as an effective theory for rapidly expanding
systems.

Excluding the contribution from other conserved quantities (such as baryon
number), the equations of motion of relativistic hydrodynamics stem from the
conservation laws of energy and momentum, $\nabla _{\mu }T^{\mu \nu }=0$,
with $T^{\mu \nu }$ being the energy-momentum tensor of the fluid. Quite
generally, one may write $T^{\mu \nu }=T_{ideal}^{\mu \nu }+\Pi ^{\mu \nu }$
with $T_{ideal}^{\mu \nu }$ being the energy-momentum tensor of an ideal
fluid constructed using the local energy density $\varepsilon $ and flow
velocity $u_{\mu }$ (i.e., the standard hydrodynamic fields) and $\Pi ^{\mu
\nu }$ being a dissipative contribution whose explicit form can only be
found with additional assumptions. In the Landau frame \cite%
{LandauLifshitzFluids} (used throughout this paper), in the absence of bulk
viscous effects $\Pi ^{\mu \nu }=\pi ^{\mu \nu }$, with $\pi ^{\mu \nu
}=\Delta _{\alpha \beta }^{\mu \nu }T^{\alpha \beta }$ being the shear
stress tensor constructed using the tensor projector $\Delta _{\alpha \beta
}^{\mu \nu }=(\Delta _{\alpha }^{\mu }\Delta _{\beta }^{\nu }+\Delta
_{\alpha }^{\nu }\Delta _{\beta }^{\mu })/2-\Delta _{\alpha \beta }\Delta
^{\mu \nu }/3$ defined by the projection operator transverse to the flow $%
\Delta _{\mu \nu }=g_{\mu \nu }-u_{\mu }u_{\nu }$ ($g_{\mu \nu }$ is the
spacetime metric). In the gradient expansion approach \cite{DeGroot:1980dk},
the dissipative fluxes, such as $\pi ^{\mu \nu }$, are organized as a formal
expansion in powers of the spacetime gradients of the hydrodynamic fields
taking into account all the possible structures compatible with the
symmetries, whose conformal limit was originally worked out in the Landau
frame to second order in gradients in \cite{Baier:2007ix,Bhattacharyya:2008jc}.

However, in the relativistic regime this approach faces considerable
challenges since the equations of motion obtained from this formalism
generally display acausal behavior and instabilities (at least at the linear
level) \cite%
{Hiscock_Lindblom_stability_1983,Hiscock_Lindblom_instability_1985} already
at first order in the gradient expansion, i.e. relativistic Navier-Stokes
theory, which are not resolved by the inclusion of second order derivatives
of the hydrodynamic fields \cite{Finazzo:2014cna} unless some type of
resummation involving the hydrodynamic fields is employed \cite%
{Bemfica:2017wps}. With the recent evidence that the gradient series has
zero radius of convergence in the relativistic regime both at strong
coupling \cite{Heller:2013fn,Buchel:2016cbj} and in kinetic theory models 
\cite{Denicol:2016bjh,Heller:2016rtz}, it is unlikely that any of these
problems are resolved perturbatively by going to even higher orders in the
expansion. This motivates the search for a meaningful definition of viscous
relativistic hydrodynamics that does not resort to an expansion in gradients
of the hydrodynamic fields.

At least in the linear regime, causality and stability can be obtained by extending
the set of dynamical variables to include not only the hydrodynamic fields
but also the dissipative fluxes, as in Israel-Stewart (IS) theory \cite%
{MIS-6}. Other approaches include, for instance, divergence type theories 
\cite{Geroch:1990bw}. In IS $\pi^{\mu\nu}$ is defined dynamically via
additional equations of motion, which were originally determined by
requiring that the second law of thermodynamics is satisfied. In this approach quantities such as $\pi^{\mu\nu}/(%
\varepsilon+P)$ (with $P=P(\varepsilon)$ being the equilibrium pressure) are
assumed to be small, though it is important to remark that this assumption
does not necessarily imply an expansion in gradients.

An interesting property displayed by IS theory and other more modern
approaches such as \cite{Denicol:2012cn} is that the first-order
relaxation-type equations obeyed by $\pi^{\mu\nu}$ imply that these
quantities must also be specified on the spacelike hypersurface that defines
the initial value problem. Since the conservation laws couple the
hydrodynamic fields to the dissipative fluxes, the solution for the
hydrodynamic fields $\{\varepsilon,u_\mu\}$ is expected to be sensitive to
the choices made for the initial conditions of $\pi^{\mu\nu}$. In
equilibrium this dependence is of course erased but one may ask whether
there is some type of non-equilibrium regime in which such a dependence is
minimal. In this regard, one may define a non-equilibrium hydrodynamic
attractor by the condition that for a large set of initial conditions%
\footnote{%
Initial conditions for the dissipative fluxes are not, in fact, completely
arbitrary. For instance, one may require the weak energy condition, $%
T_{\mu\nu}t^\mu t^\nu \geq 0$ where $t^\mu$ is an arbitrary time-like
4-vector \cite{HawkingEllisBook}, to be satisfied. For a discussion on
related topics, see \cite{Arnold:2014jva}.} the system's dynamics collapses
at large times onto a single encompassing behavior, before true thermal
equilibrium is reached.

This feature was observed \cite{Heller:2011ju} in a numerical simulation of
strongly coupled $\mathcal{N}=4$ Supersymmetric Yang-Mills (SYM) theory with
large number of colors undergoing Bjorken flow \cite{Bjorken:1982qr} and its
meaning was clarified in \cite{Heller:2015dha} via a study of conformal IS
theory also assuming Bjorken symmetry. Since then, such dynamical attractor
behavior has been investigated in other works \cite%
{Romatschke:2017vte,Florkowski:2017olj,Spalinski:2017mel,Bemfica:2017wps,Strickland:2017kux,Romatschke:2017acs,Florkowski:2017jnz}%
. The presence of an attractor solution restores the large degree of
universality usually associated with hydrodynamic behavior without relying
on the gradient expansion.

In Bjorken expanding systems the symmetries are so powerfully constraining
that it is possible to investigate the large order limit of the gradient
expansion in a systematic manner \cite{Heller:2013fn} (the same holds for
fluids embedded in an expanding Universe \cite{Buchel:2016cbj}), which is
not feasible in less symmetric situations. This allowed the authors of Ref.\ 
\cite{Heller:2015dha} to show that the hydrodynamic attractor corresponds to
a resummation of the gradient series, establishing an interesting link
between hydrodynamics and the mathematics of resurgence theory, later
pursued by other works \cite%
{Basar:2015ava,Aniceto:2015mto,Buchel:2016cbj,Spalinski:2017mel}.

The numerically obtained attractor solutions found so far indicate that it
is possible to find universal hydrodynamic behavior far-from-equilibrium,
regardless of the details of the initial state of the system. However, even
though one may now associate hydrodynamic behavior with such non-equilibrium
attractors, it is not straightforward to clearly state its domain of
validity or even how to clearly define attractors (besides by explicit
numerical inspection involving a large number of initial conditions).

Given the simplicity of the hydrodynamic equations in Bjorken flow, in this
case it is possible to use different ways to identify the non-equilibrium
attractor, as discussed in \cite{Heller:2015dha}. One method involves
defining the boundary condition of the fields at very early times. Another
possibility is the resummation of the divergent gradient series. The last
method is the analog of the slow-roll expansion used in cosmology \cite%
{Liddle:1994dx} whose zeroth order term already generally gives a decent
approximation for the attractor in the Bjorken case. Further progress in
identifying the virtues and issues with these approaches could be achieved
by having an analytical example where the fluid's evolution towards the
attractor can be investigated in a simpler way.

In this paper we show that the equations of viscous relativistic
hydrodynamics (neglecting effects from bulk viscosity and chemical
potential) can be solved analytically when the shear relaxation time is
constant and the system undergoes a Bjorken expansion. Differently than
other studies, here we analytically determine the hydrodynamic attractor of
this system in closed form and discuss its properties. We perform the first
study of the large order behavior of the slow-roll expansion and compare it
to the analytical attractor. We find that the slow-roll expansion in
hydrodynamics diverges. This is also the case for conformal fluids. We
investigate the role played by the values of the transport coefficients on
the convergence of the gradient expansion and show that the series can
actually converge if the transport coefficients do not fulfill the standard
conditions for causality and stability determined from well-known linearized
analyses \cite{Hiscock_Lindblom_stability_1983} (see also \cite{Pu:2009fj}). We also discuss the generalized gradient expansion series first presented in \cite{Denicol:2016bjh} and we apply it here to find solutions of IS theory. In contrast to the other series discussed, this one appears to converge and offers a very
good description of the analytical solution already at second order.

This paper is organized as follows. In the next section we define the
viscous hydrodynamic equations we use and obtain their full analytical
solution, under the conditions mentioned above. We analytically determine in
Section \ref{sec:attractor} the non-equilibrium attractor and discuss
several of its properties. Our conclusions and outlook are presented in \ref%
{sec:conclusions}.

\emph{Definitions:} Throughout this work we use natural units $\hbar=c=k_B=1$
and Milne coordinates where $x^\mu = (\tau,x,y,\varsigma)$, with proper-time
and spacetime rapidity defined in terms of standard Minkowski coordinates as
follows: $\tau = \sqrt{t^2-z^2}$ and $\varsigma = \tanh^{-1}(z/t)$. In these
coordinates, the metric is $ds^2 = g_{\mu\nu}dx^\mu dx^\nu=d\tau^2 - dx^2
-dy^2 -\tau^2 d\varsigma^2$, and the corresponding nonzero Christoffel
symbols are $\Gamma_{\varsigma}^{\tau\varsigma} =
\Gamma_{\varsigma\tau}^\varsigma = 1/\tau$ and $\Gamma_{\varsigma\varsigma}^%
\tau = \tau$.

\section{Analytical solution of viscous relativistic hydrodynamics in
Bjorken flow}

\label{sec:anal}

The set of viscous relativistic hydrodynamic equations we use is given by 
\begin{eqnarray}
&&D\varepsilon +(\varepsilon +P)\theta -\pi ^{\mu \nu }\sigma _{\mu \nu }=0
\\
&&(\varepsilon +P)Du^{\mu }-\Delta _{\lambda }^{\mu }\nabla ^{\lambda
}P+\Delta _{\lambda }^{\mu }\nabla _{\mu }\pi ^{\mu \lambda }=0 \\
&&\tau _{R}\Delta _{\alpha \beta }^{\mu \nu }D\pi ^{\alpha \beta }+\delta
_{\pi \pi }\,\theta \pi ^{\mu \nu }+\tau _{\pi \pi }\,\Delta _{\alpha \beta
}^{\mu \nu }\pi ^{\alpha \lambda }\sigma _{\lambda }^{\beta }-2\,\tau
_{R}\Delta _{\alpha \beta }^{\mu \nu }\pi _{\lambda }^{\alpha }\omega
^{\beta \lambda }+\pi ^{\mu \nu }=2\eta \sigma ^{\mu \nu },  \label{eqsDNMR}
\end{eqnarray}%
where $D=u^{\mu }\nabla _{\mu }$ is the co-moving covariant derivative, $%
\theta =\nabla _{\mu }u^{\mu }$ is the local expansion rate, $\sigma _{\mu
\nu }=\Delta _{\mu \nu }^{\alpha \beta }\nabla _{\alpha }u_{\beta }$ is the
shear tensor, $\omega _{\mu \nu }=(\Delta _{\mu }^{\lambda }\nabla _{\lambda
}u_{\nu }-\Delta _{\nu }^{\lambda }\nabla _{\lambda }u_{\mu })/2$ is the
vorticity tensor, $\eta $ is the shear viscosity, and $\tau _{R}$ is the
shear relaxation time. We neglect all effects from bulk viscous pressure and
assume an ideal gas equation of state, $\varepsilon =3P$, at zero chemical
potential. The equations above may be derived using the Boltzmann equation
in the 14-moment approximation or in the relaxation time approximation
(RTA), as shown in Refs.~\cite{Denicol:2012cn,Denicol:2012es,Denicol:2014vaa}%
. In the 14-moment approximation and for a massless gas, one can show that $%
\delta _{\pi \pi }=4/3\,\tau _{R}$, $\tau _{\pi \pi }=10/21\,\tau _{R}$ and $%
\eta =(\varepsilon +P)\tau _{R}/5$ \cite%
{Denicol:2012cn,Denicol:2012es,Denicol:2014vaa}. For now, we assume a more
general expression for $\tau _{\pi \pi }$, where $\tau _{\pi \pi }=\lambda
\,\tau _{R}$. In this paper we further assume that $\tau _{R}$ is constant,
an assumption that plays a crucial role in determining the analytical
solution derived below.

We impose Bjorken symmetry and, thus, in our coordinate system $u_{\mu }=(1,0,0,0)$. This implies
that only the first and the third equations above contain nontrivial
information. The symmetries further constrain the expansion rate of the
fluid, $\theta =1/\tau $, and its shear tensor, $\sigma _{\mu \nu }$, which
becomes diagonal%
\begin{equation}
\sigma _{\nu }^{\mu }=\mathrm{diag}\left( 0,-\frac{1}{3\tau },-\frac{1}{%
3\tau },\frac{2}{3\tau }\right) .
\end{equation}%
Furthermore, the shear stress tensor will also become diagonal and can be
described with only one independent degree of freedom,

\begin{equation}
\pi _{\nu }^{\mu }=\mathrm{diag}(0,-\pi /2,-\pi /2,\pi ).
\end{equation}%
Therefore, even though homogeneous, the system is not static and, at
sufficiently early times, the gradients $\theta $ and $\sigma ^{\mu \nu }$
can become large enough to drive the system far away from local
thermodynamic equilibrium. On the other hand, the gradients of any scalar,
such as the chemical potential and temperature, are always zero, prohibiting
the existence of any heat flow or diffusion. Finally, in this geometry the
vorticity tensor is also always zero%
\begin{equation}
\omega ^{\mu \nu }=0.
\end{equation}%
The evolution of the fluid is then described by the following set of coupled
differential equations, 
\begin{eqnarray}
&&\frac{d\varepsilon }{d\tau }+\frac{(\varepsilon +P)}{\tau }-\frac{\pi }{%
\tau }=0, \\
&&\tau _{R}\frac{d\pi }{d\tau }+\pi +\left( \frac{4}{3}+\lambda \right) \tau
_{R}\frac{\pi }{\tau }=\frac{4}{3}\frac{\eta }{\tau }.  \label{definemodel1}
\end{eqnarray}

It is convenient to rewrite these equations in terms of the dimensionless
field, $\chi \equiv \pi /(\varepsilon +P)$, and define the dimensionless
propertime variable $\hat{\tau}=\tau /\tau _{R}$, which is the inverse Knudsen, $K_{N}=1/\hat{\tau}$, in Bjorken flow \cite{Denicol:2016bjh}. With these changes of variables, the
equations simplify to 
\begin{equation}
\frac{1}{\varepsilon \hat{\tau}^{4/3}}\frac{d(\varepsilon \hat{\tau}^{4/3})}{%
d\hat{\tau}}=\frac{4}{3}\frac{\chi }{\hat{\tau}},  \label{eqe}
\end{equation}%
and 
\begin{equation}
\frac{d\chi }{d\hat{\tau}}+\lambda \frac{\chi }{\hat{\tau}}+\frac{4}{3\hat{%
\tau}}\chi ^{2}+\chi -\frac{3}{4}\frac{a}{\hat{\tau}}=0,  \label{eqpi}
\end{equation}%
with 
\begin{equation}
a=\frac{16}{9(\tau _{R}T)}\frac{\eta }{s}.
\end{equation}%
Causality and stability around equilibrium at the linearized level hold when 
$\eta /(s\tau _{R}T)\leq 1/2$ \cite{Pu:2009fj}, i.e., $a\leq 8/9$. Even
though $a=16/45$ and $\lambda =10/21$ in the 14-moment approximation, we
kept $a$ and $\lambda $ arbitrary above since the general analytical
solution of these equations can be found for any $a\geq 0$ and $\lambda \in 
\mathbb{R}$, as we show below.

Equation\ \eqref{eqpi} is a Riccati equation that can be solved
independently of \eqref{eqe}, a direct consequence of the constant $\tau _{R}
$ assumption. First order nonlinear ODEs of Riccati type can always be
written as second order \emph{linear} ODE's and, as a matter of fact, this
can be done in the present case using a new variable $y(\hat{\tau})$ defined
via 
\begin{equation}
\frac{1}{y}\frac{dy}{d\hat{\tau}}=\frac{4}{3}\frac{\chi }{\hat{\tau}}.
\label{definey}
\end{equation}%
Inserting this into \eqref{eqpi}, provides 
\begin{equation}
\frac{d^{2}y}{d\hat{\tau}^{2}}+\left( 1+\frac{1+\lambda }{\hat{\tau}}\right) 
\frac{dy}{d\hat{\tau}}-\frac{a}{\hat{\tau}^{2}}y=0.  \label{neweqpi}
\end{equation}%
This linear ODE can be solved and the general solution is 
\begin{equation}
y(\hat{\tau})=A\,\hat{\tau}^{-(\lambda +1)/2}\exp (-\hat{\tau}/2)\left[ M_{-%
\frac{\lambda +1}{2},\frac{\sqrt{\lambda ^{2}+4a}}{2}}(\hat{\tau})+\alpha
\,W_{-\frac{\lambda +1}{2},\frac{\sqrt{\lambda ^{2}+4a}}{2}}(\hat{\tau})%
\right] ,  \label{solutiony}
\end{equation}%
where $A$ and $\alpha $ are constants and $M_{k,\mu }(z)$ and $W_{k,\mu }(z)$
are Whittaker functions\footnote{%
We refer the reader to Ref.\ \cite{abramowitzstegun} for more details
concerning the analytical structure of these functions.}. Using %
\eqref{solutiony}, one can find the following analytical solution for the
energy density 
\begin{eqnarray}
\varepsilon (\hat{\tau}) &=&\varepsilon _{0}\left( \frac{\hat{\tau}_{0}}{%
\hat{\tau}}\right) ^{4/3}\frac{y(\hat{\tau})}{y(\hat{\tau}_{0})}  \notag
\label{analenergy} \\
&=&\varepsilon _{0}\left( \frac{\hat{\tau}_{0}}{\hat{\tau}}\right) ^{\frac{4%
}{3}+\frac{\lambda +1}{2}}\,\exp \left( -\frac{\hat{\tau}-\hat{\tau}_{0}}{2}%
\right) \left[ \frac{M_{-\frac{\lambda +1}{2},\frac{\sqrt{\lambda ^{2}+4a}}{2%
}}(\hat{\tau})+\alpha \,W_{-\frac{\lambda +1}{2},\frac{\sqrt{\lambda ^{2}+4a}%
}{2}}(\hat{\tau})}{M_{-\frac{\lambda +1}{2},\frac{\sqrt{\lambda ^{2}+4a}}{2}%
}(\hat{\tau}_{0})+\alpha \,W_{-\frac{\lambda +1}{2},\frac{\sqrt{\lambda
^{2}+4a}}{2}}(\hat{\tau}_{0})}\right]   \notag \\
&&
\end{eqnarray}%
and the normalized shear stress tensor component 
\begin{equation}
\chi (\hat{\tau})=\frac{\pi }{\varepsilon +P}=\frac{3\left( \sqrt{4a+\lambda
^{2}}-\lambda \right) M_{\frac{1}{2}-\frac{\lambda }{2},\frac{1}{2}\sqrt{%
4a+\lambda ^{2}}}\left( \hat{\tau}\right) -6\alpha W_{\frac{1}{2}-\frac{%
\lambda }{2},\frac{1}{2}\sqrt{4a+\lambda ^{2}}}\left( \hat{\tau}\right) }{%
8\left( M_{-\frac{1+\lambda }{2},\frac{1}{2}\sqrt{4a+\lambda ^{2}}}\left( 
\hat{\tau}\right) +\alpha W_{-\frac{1+\lambda }{2},\frac{1}{2}\sqrt{%
4a+\lambda ^{2}}}\left( \hat{\tau}\right) \right) }.  \label{analpibar}
\end{equation}%
One can see that the value of $A$ in \eqref{solutiony} does not enter in
either $\varepsilon $ or $\pi $. Thus, the constants that define the
initial-value problem at $\hat{\tau}=\hat{\tau}_{0}>0$ are $\varepsilon _{0}$
and $\alpha $, since the latter can be written in terms of $\chi (\hat{\tau}%
_{0})$. One important constraint for this solution is that $\alpha $ must be
such that $y(\hat{\tau})$ remains non-negative for all $\hat{\tau}\geq \hat{%
\tau}_{0}$ to make sure that the energy density is positive-definite and
there are no zeros in the denominators of the expressions above. Equations\ %
\eqref{analenergy} and \eqref{analpibar} define the general analytical
solution of the viscous hydrodynamic equations in Bjorken flow with a
constant relaxation time. As such, they can be easily implemented in studies
of different hydrodynamic schemes and their comparison to exact solutions in
kinetic theory, such as \cite{Florkowski:2013lya}.

\section{Analytical non-equilibrium attractor}

\label{sec:attractor}

In this section we investigate the solution of the hydrodynamic equations
and the corresponding non-equilibrium attractor. No significant change is
observed when $\lambda $ is taken into account and, thus, we set $\lambda =0$
in the following (this approximation was also used in \cite{Heller:2015dha}%
). For convenience, we repeat the equation for $\chi $ in this case below 
\begin{equation}
\hat{\tau}\frac{d\chi }{d\hat{\tau}}+\frac{4}{3}\chi ^{2}+\hat{\tau}\chi -%
\frac{3a}{4}=0.  \label{eqphi}
\end{equation}%
The general analytical solution of this equation can also be written in
terms of Bessel functions 
\begin{equation}
\chi (\hat{\tau})=\frac{3\sqrt{a}}{4}\left[ \frac{\alpha \left( K_{\sqrt{a}-%
\frac{1}{2}}\left( \frac{\hat{\tau}}{2}\right) +K_{\sqrt{a}+\frac{1}{2}%
}\left( \frac{\hat{\tau}}{2}\right) \right) +I_{\sqrt{a}-\frac{1}{2}}\left( 
\frac{\hat{\tau}}{2}\right) -I_{\sqrt{a}+\frac{1}{2}}\left( \frac{\hat{\tau}%
}{2}\right) }{\alpha \left( K_{\sqrt{a}-\frac{1}{2}}\left( \frac{\hat{\tau}}{%
2}\right) -K_{\sqrt{a}+\frac{1}{2}}\left( \frac{\hat{\tau}}{2}\right)
\right) +I_{\sqrt{a}-\frac{1}{2}}\left( \frac{\hat{\tau}}{2}\right) +I_{%
\sqrt{a}+\frac{1}{2}}\left( \frac{\hat{\tau}}{2}\right) }\right] 
\label{analpi1}
\end{equation}%
and the corresponding expression for the energy density is 
\begin{equation}
\varepsilon (\hat{\tau})=\varepsilon _{0}\,e^{-\frac{1}{2}\left( \hat{\tau}-%
\hat{\tau}_{0}\right) }\left( \frac{\hat{\tau}_{0}}{\hat{\tau}}\right) ^{%
\frac{5}{6}}\left[ \frac{\alpha \left( K_{\sqrt{a}-\frac{1}{2}}\left( \frac{%
\hat{\tau}}{2}\right) -K_{\frac{1}{2}+\sqrt{a}}\left( \frac{\hat{\tau}}{2}%
\right) \right) +I_{\sqrt{a}-\frac{1}{2}}\left( \frac{\hat{\tau}}{2}\right)
+I_{\frac{1}{2}+\sqrt{a}}\left( \frac{\hat{\tau}}{2}\right) }{\alpha \left(
K_{\sqrt{a}-\frac{1}{2}}\left( \frac{\hat{\tau}_{0}}{2}\right) -K_{\frac{1}{2%
}+\sqrt{a}}\left( \frac{\hat{\tau}_{0}}{2}\right) \right) +I_{\sqrt{a}-\frac{%
1}{2}}\left( \frac{\hat{\tau}_{0}}{2}\right) +I_{\frac{1}{2}+\sqrt{a}}\left( 
\frac{\hat{\tau}_{0}}{2}\right) }\right] .  \label{analenergy1}
\end{equation}%
We note that $\alpha \leq 0$, which guarantees that the denominator of the
above equations is always nonzero. Also, the general solution \eqref{analpi1} cannot be simply decomposed in terms of an attractor plus transient corrections. Both contributions are present in the numerator and the denominator of the analytical solution.   

In all the previous studies on non-equilibrium attractors in Bjorken flow
the attractor per se was only found numerically using basically three
different approaches \cite{Heller:2015dha}:

\begin{itemize}
\item Explicit construction by solving the corresponding differential
equation fixing a specific boundary condition at very early times.

\item Resummation of the gradient series.

\item Slow-roll expansion.
\end{itemize}

We will use the analytical solution found here to illustrate how these
approaches fare at identifying the analytical attractor and its properties.

\subsection{The attractor solution}

\begin{figure}[th]
\includegraphics[width=0.5\textwidth]{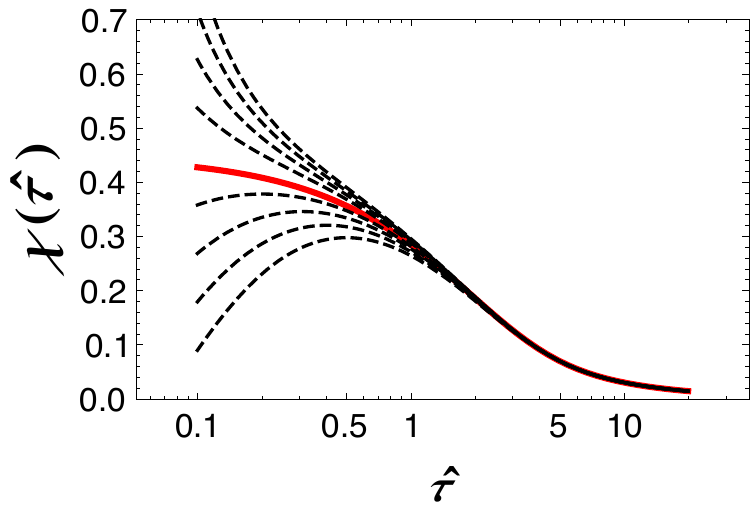}
\caption{(color online) Analytical non-equilibrium attractor 
\eqref{analpi1attractor} in solid red for the RTA value $a=16/45$ (and $%
\protect\lambda =0$). The dashed curves correspond to the solution in Eq.\ 
\eqref{analpi1} for different initial conditions (parametrized by $\protect%
\alpha $), which collapse onto the attractor before local equilibrium is
reached (where $\protect\chi $ vanishes).}
\label{fig:attractor}
\end{figure}

From the analytical solution derived in the previous section, it is
straightforward to see that the solution \eqref{analpi1} completely loses the
information about the initial conditions (encoded in $\alpha $) at late
times. This happens because the Bessel functions display the following
asymptotic form for sufficiently large values of its argument, $K_{\nu
}(x)\sim e^{-x}/\sqrt{x}$ and $I_{\nu }(x)\sim e^{x}/\sqrt{x}$. Therefore,
the terms containing $K_{\nu }\left( \hat{\tau}/2\right) $ become
significantly smaller compared to the terms containing $I_{\nu }\left( \hat{%
\tau}/2\right) $ as time increases. At a sufficiently long time, the
solution \eqref{analpi1} can be approximated as 
\begin{equation}
\chi (\hat{\tau})\rightarrow \chi _{att}(\hat{\tau})=\frac{3\sqrt{a}}{4}%
\left[ \frac{I_{\sqrt{a}-\frac{1}{2}}\left( \frac{\hat{\tau}}{2}\right) -I_{%
\sqrt{a}+\frac{1}{2}}\left( \frac{\hat{\tau}}{2}\right) }{I_{\sqrt{a}-\frac{1%
}{2}}\left( \frac{\hat{\tau}}{2}\right) +I_{\sqrt{a}+\frac{1}{2}}\left( 
\frac{\hat{\tau}}{2}\right) }\right] .  \label{analpi1attractor}
\end{equation}%
This expression corresponds to the exact solution for $\alpha =0$ and it
represents the non-equilibrium attractor solution of the hydrodynamic
equations investigated here. The typical attractor behavior is illustrated
in Fig.\ \ref{fig:attractor} for the RTA case where $a=16/45$. One can see
that \eqref{analpi1attractor} is the only solution of the differential
equation that smoothly connects to the positive $\chi $ branch at early
times, i.e., $\lim_{\hat{\tau}\rightarrow 0}\chi (\hat{\tau})\Big |_{a=\frac{%
16}{45}}=1/\sqrt{5}$. 

When discussing the attractor solution in Bjorken flow, a common procedure consists in analyzing the behavior of $\chi$ at $\hat\tau \to 0 $. In our case this gives 
 two limiting values: $3\sqrt{a}/4$ for $\alpha=0$ and $-3\sqrt{a}/4$ if $\alpha \neq 0$. Therefore, the attractor is the only solution that goes to $3\sqrt{a}/4$ at $\htau=0$. Indeed, this limiting behavior of the attractor in Bjorken flow has been used in previous works as a way to define it \cite{Heller:2015dha}. In this case, one may find the attractor numerically by identifying it as the solution that obeys this boundary condition. 
 
\subsection{Resummation of the gradient series}

The formal gradient expansion solution is represented as the late time
series $\chi (\hat{\tau})=(3a/4)\sum_{n=0}^{\infty }c_{n}/\hat{\tau}^{n}$,
where the corresponding coefficients of the series are given by 
\begin{equation}
c_{n+1}=n\,c_{n}-a\sum_{m=0}^{n}c_{n-m}c_{m},  \label{eqdiverge}
\end{equation}%
with $c_{0}=0$ and $c_{1}=1$. It is interesting to notice that when
causality and stability are fulfilled, i.e. for $a\leq 8/9$, the gradient
series diverges since for large $n$ the first term in \eqref{eqdiverge}
dominates leading to factorial growth.

Setting $a=1$ is particularly interesting since in this case one can show
that all $c_{n\geq 1}=1$, which leads to 
\begin{equation}
\chi (\hat{\tau})=\frac{3}{4}\sum_{n=1}^{\infty }\frac{1}{\hat{\tau}^{n}}.
\label{seriesa}
\end{equation}%
In contrast to the other examples in Bjorken flow, this series has a nonzero
radius of convergence, i.e., for any $\hat{\tau}>1$ (which overlaps with the
expected domain of the late time series) this can be summed up to give 
\begin{equation}
\chi (\hat{\tau})\rightarrow \frac{3}{4(\hat{\tau}-1)}.
\end{equation}%
However, we remark that this nonzero radius of convergence was possible only
when $a$ was taken in the acausal region. Convergent series can also be
obtained for other values of $a$, e.g., $a=4$ and 9. However, all these
values are in the acausal regime.

Now, to see that resumming the gradient series does lead to the attractor,
we note that the full analytical solution in the case $a=1$ is 
\begin{equation}
\chi (\hat{\tau})\Big|_{a=1}=\frac{3}{4}\left[ \frac{\frac{1}{\hat{\tau}}%
+\alpha \left( 1+\frac{1}{\hat{\tau}}\right) e^{-\hat{\tau}}}{1-\frac{1}{%
\hat{\tau}}+\frac{\alpha }{\hat{\tau}}e^{-\hat{\tau}}}\right] 
\label{generala1}
\end{equation}%
and correspondingly the attractor solution \eqref{analpi1attractor} is
simply 
\begin{equation}
\chi (\hat{\tau})\Big|_{a=1,att}=\frac{3}{4\left( \hat{\tau}-1\right) },
\label{atta1}
\end{equation}%
which matches the result obtained from the gradient series. Even though
there is a pole in $\hat{\tau}=1$, we note that the expression above is
meaningful for larger times.

The current example with $a=1$ shows in a very clear manner that the
attractor can be defined via a resummation of the gradient series.
Furthermore, it is straightforward to find a late time trans-series
representation for the general analytical solution in \eqref{generala1}. The
first terms are 
\begin{eqnarray}
\chi(\hat\tau) &=& \frac{3}{4}\sum_{n=0}^\infty \frac{1}{\hat\tau^{n+1}} - 
\frac{3}{4}\alpha\, e^{-\hat\tau}\sum_{n=0}^{\infty} \frac{1}{\hat\tau^n}%
\left(1+\frac{1}{\hat\tau}+\frac{n+1}{\hat\tau^{2}}\right)+\mathcal{O}%
(\alpha^2 e^{-2 \hat\tau}) \\
&=& \frac{3}{4}\frac{1}{(\hat\tau-1)}-\frac{3}{4}\alpha\,e^{-\hat\tau} \frac{%
\hat\tau^2}{(\hat\tau-1)^2}+\mathcal{O}(\alpha^2 e^{-2 \hat\tau}),
\end{eqnarray}
where one can see that $\alpha$, the parameter that defines the initial
condition, plays the role of the trans-series expansion parameter \cite%
{Basar:2015ava}. In this case, the contribution from each term in the
trans-series can be easily determined since their corresponding power series
representation converge (no Borel transforms are needed). For the causal
configuration where $a\leq 8/9$, this is not the case and one must resort to
resurgence theory to resum the series. In this case, one can compare the
result from the resummation directly to the analytical expression for the
attractor, which may lead to further insight into the application of
resurgence ideas in hydrodynamics. However, this is beyond the scope of the
present paper and we leave this interesting task for a future study.

\subsection{Divergence of the slow-roll expansion}

\begin{figure}[th]
\includegraphics[width=0.6\textwidth]{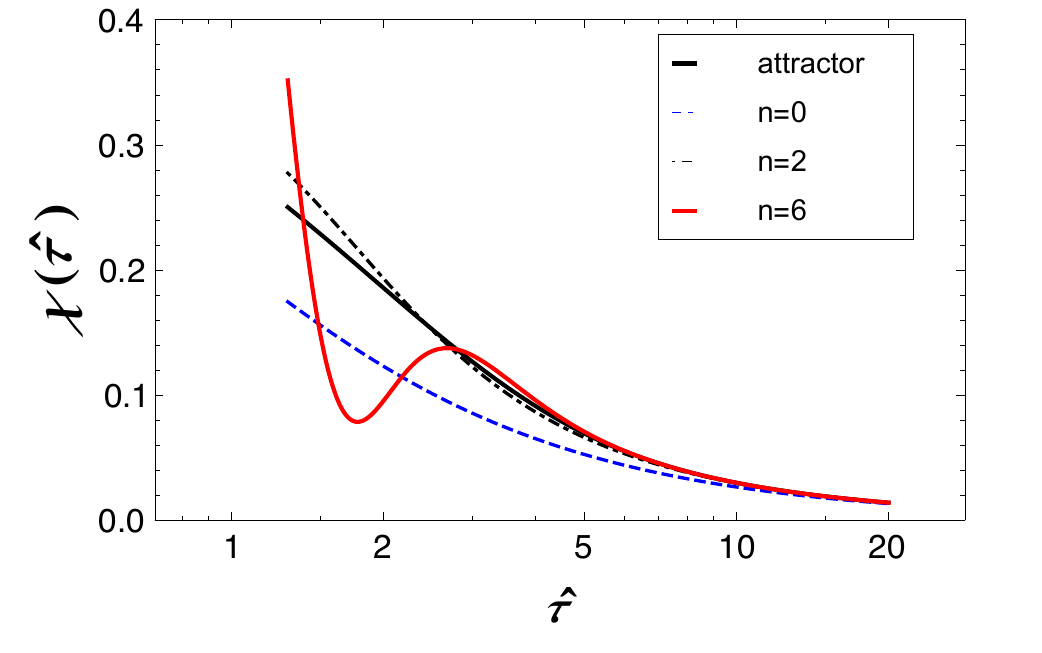}
\caption{(color online) Comparison between the analytical attractor in 
\eqref{atta1} for $a=16/45$ and the result from the slow-roll expansion,
computed at different orders.}
\label{fig:compatt}
\end{figure}

\begin{figure}[th]
\includegraphics[width=0.6\textwidth]{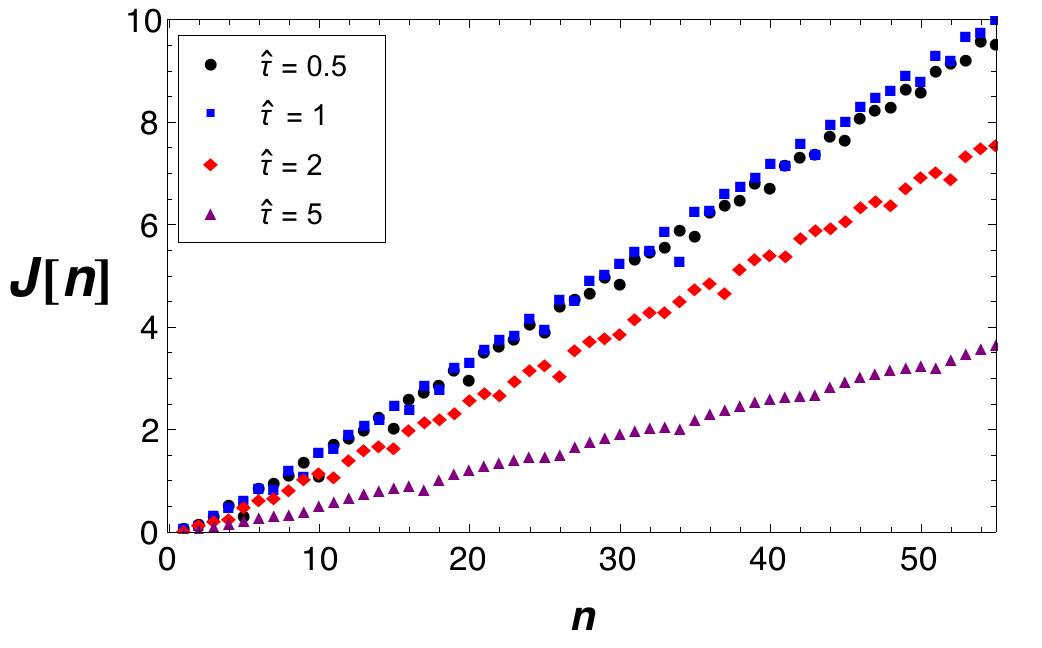}
\caption{(color online) Large order behavior of the slow-roll expansion in hydrodynamics. $J[n]=|\chi_n|^{1/n}$ as a function of $n$ for $a=16/45$ and different times $\hat{\protect\tau}=0.5,1,2,5
$. }
\label{fig:divergeslowroll}
\end{figure}

Reference\ \cite{Heller:2015dha} suggested another way to characterize the
attractor based on the analog of the slow-roll expansion used in cosmology 
\cite{Liddle:1994dx}. This can be done systematically \cite%
{Strickland:2017kux} by including a small parameter $\epsilon $ (not to be
confused with the energy density $\varepsilon $) in the differential
equation 
\begin{equation}
\epsilon \,\hat{\tau}\frac{d\chi }{d\hat{\tau}}+\frac{4}{3}\chi ^{2}+\hat{%
\tau}\chi -\frac{3a}{4}=0  \label{eqphislowroll}
\end{equation}%
where now $\chi =\chi (\hat{\tau};\epsilon )$ is also a function of $%
\epsilon $. The next step is to look for a series solution in powers of $%
\epsilon $ for $\chi $ 
\begin{equation}
\chi (\hat{\tau};\epsilon )=\sum_{n=0}^{\infty }\chi _{n}(\hat{\tau}%
)\epsilon ^{n}.
\end{equation}%
Clearly, the full answer is only obtained in the limit $\epsilon \rightarrow
1$. In practice, $\epsilon $ is taken to 1 already after including only a
few terms, given the apparent convergence of this procedure found in
previous works. The zeroth order term gives two solutions and the one that
recovers the Navier-Stokes (NS) limit at late times, $\chi _{NS}\sim 3a/(4%
\hat{\tau})$, is 
\begin{equation}
\chi _{0}(\hat{\tau})=\frac{3}{8}\left( \sqrt{\hat{\tau}^{2}+4a}-\hat{\tau}%
\right) .
\end{equation}%
The other terms with $n\geq 1$ can be found from the recurrence relation 
\begin{equation}
\chi _{n}(\hat{\tau})=-\frac{1}{\sqrt{\hat{\tau}^{2}+4a}}\left( \hat{\tau}%
\frac{d\chi _{n-1}}{d\hat{\tau}}+\frac{4}{3}\sum_{m=1}^{n-1}\chi _{n-m}\chi
_{m}\right) .
\end{equation}%
Each term of the series can be determined analytically, which may be used to
study the large order behavior of the slow-roll expansion in hydrodynamics.
We show in Fig.\ \ref{fig:compatt} a comparison between the analytical
attractor in \eqref{atta1} (solid black curve) and the result from the
slow-roll expansion computed at different orders for $a=16/45$. One can see
that there is an improvement when going from 0 to 2rd order as the latter
gives a good representation for the attractor for $\hat{\tau}\geq 3$.
However, as we increase the order of the expansion, already at $n=6$ the
result oscillates significantly, which indicates that the slow-roll
expansion does not converge. 
In fact, this is indeed the case as shown in Fig.\ \ref{fig:divergeslowroll}
which shows for the first time the large order behavior of the slow-roll
expansion in hydrodynamics. One can see that for different values of $\hat{%
\tau}$ the series appears to diverge. This behavior persisted for all values
of $a$ in the causal regime\footnote{%
The maximum number of terms we could investigate numerically goes to roughly
100. Going to larger times also did not change this behavior. We also
checked values of $a$ in the acausal regime. Again, no qualitative
difference was found.}. Therefore, both the gradient series and the
slow-roll expansion diverge in hydrodynamics. 
However, this divergence does not mean that such series are not useful. As a
matter of fact, when properly truncated divergent series provide extremely
powerful approximations to the solutions of several problems \cite{bender}.

\begin{figure}[th]
\begin{center}
\includegraphics[width=0.6\textwidth]{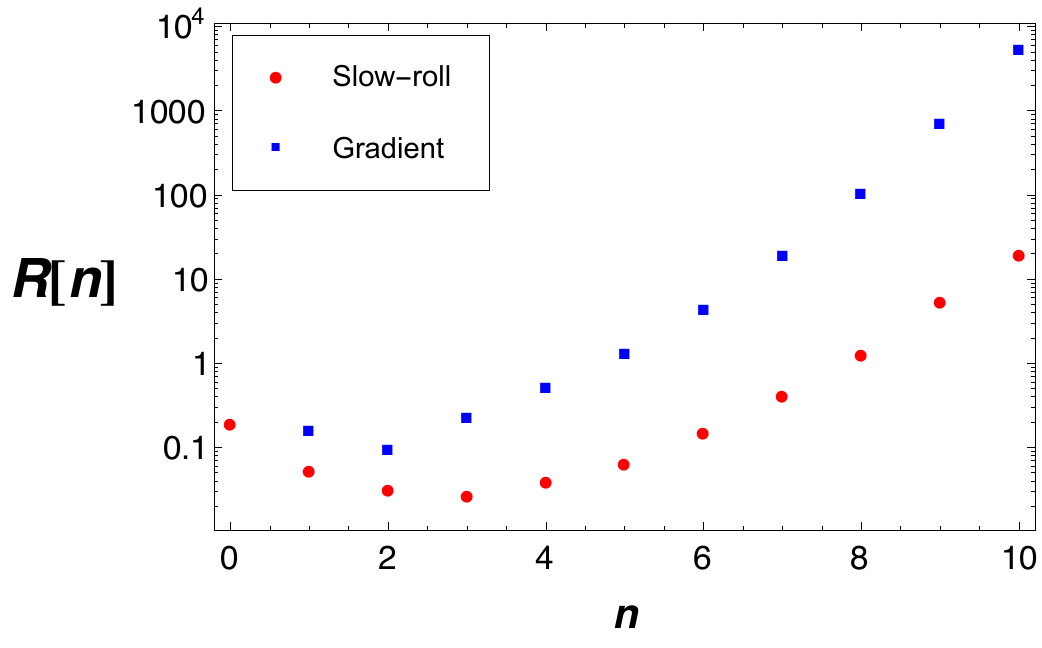}
\end{center}
\caption{(color online) Comparison between the relative difference between
the attractor and the gradient and slow-roll expansions defined in 
\eqref{diff} as a function of the truncation order $n$ (with $a=16/45$).}
\label{fig:integralratioabs}
\end{figure}

To illustrate that this is the case here, we plot the relative difference
between the attractor and the two different series representations. In Fig.\ %
\ref{fig:integralratioabs} we plot 
\begin{equation}
R[n]=\frac{\int_{\hat{\tau}_{0}}^{\hat{\tau}_{f}}d\hat{\tau}\,\Big|\chi
_{att}(\hat{\tau})-\sum_{m=0}^{n}\chi _{n}(\hat{\tau})\Big|}{\int_{\hat{\tau}%
_{0}}^{\hat{\tau}_{f}}d\hat{\tau}\,\chi _{att}(\hat{\tau})}  \label{diff}
\end{equation}%
for the slow-roll expansion and also the corresponding expression for the
gradient series ($\hat{\tau}_{0}=1$ and $\hat{\tau}_{f}=50$). This quantity
is defined in a way that maximizes the differences between these functions
and the attractor. We see that $n=2$ seems to be the optimal truncation for
the gradient series while for the slow-roll expansion one finds $n=3$.
Altogether, the slow-roll expansion provides a much more accurate
description of the attractor than the gradient series does at any order in
the truncation (this is still the case when larger values of $n$ are
considered). However, for large values of $n$, the description already
becomes very poor. Nevertheless, we note that the truncated slow-roll series
is found to oscillate \emph{around} the attractor while the gradient series
completely misses the behavior of the attractor, leading to very different
divergence patterns. 

Therefore, the gradient expansion and the slow-roll series cannot be
used to systematically approximate the hydrodynamic solution via the inclusion of higher
order contributions. Nevertheless, the optimal truncation of these series
can be extremely useful as they provide excellent approximations for the
solution of the equations in the attractor regime.

\subsection{Divergence of the slow-roll expansion in conformal hydrodynamics}

To show that the divergence of the slow-roll expansion is not particular to
the model studied here, in this section we determine the large order
behavior of this series also in conformal hydrodynamics. In this case, the
equations of motion are still given by \eqref{definemodel1} but now $\tau
_{R}=c_{R}/T$, with $c_{R}$ being a constant. We still assume $\lambda =0$
for simplicity. We also note that in contrast with the previous case
involving a constant relaxation time, in a conformal fluid $c_{\eta }=\eta /s
$ is constant and, for instance, for a massless gas within the 14-moment
approximation $c_{R}=5\eta /s$ \cite%
{Denicol:2010xn,Denicol:2011fa,Denicol:2012cn}.

The equation for the energy density in the conformal fluid is still the same
as \eqref{eqe} but the corresponding equation for normalized shear stress
tensor component is 
\begin{equation}
c_{R}\tau \frac{d\chi }{d\tau }+\frac{4c_{R}}{3}\chi ^{2}+\chi (\tau T)-%
\frac{4}{3}c_{\eta }=0.  \label{eqpicft}
\end{equation}%
We now follow \cite{Heller:2015dha} and define the variable $w=\tau T$ (the
reciprocal of the Knudsen number for this conformal fluid), with which one
can eliminate $T$ from the equation above and find a single equation that
determines the state of the fluid 
\begin{equation}
\frac{\bar{w}}{3}\left( \chi +2\right) \chi ^{\prime }+\frac{4}{3}\chi
^{2}+\chi \bar{w}-\frac{4}{3}c_{\eta /R}=0,
\end{equation}%
where $\bar{w}=w/c_{R}$, $c_{\eta /R}=c_{\eta }/c_{R}$ \cite%
{Strickland:2017kux}, and $\chi ^{\prime }=d\chi /d\bar{w}$. We follow the
same procedure as before to obtain the slow-roll expansion $\chi (\bar{w}%
)=\sum_{n=0}^{\infty }\epsilon ^{n}\chi _{n}(\bar{w})$. The zeroth order term
that recovers the NS limit is 
\begin{equation}
\chi _{0}(\bar{w})=\frac{1}{8}\left( \sqrt{9\bar{w}^{2}+64c_{\eta /R}}-3\bar{%
w}\right) 
\end{equation}%
while the higher order terms are given by 
\begin{equation}
\chi _{n}(\bar{w})=-\frac{1}{\sqrt{9\bar{w}^{2}+64c_{\eta /R}}}\left[ 2\bar{w%
}\chi _{n-1}^{\prime }+\bar{w}\chi _{n-1}\chi _{0}^{\prime
}+\sum_{m=1}^{n-1}\left( 4\chi _{n-m}\chi _{m}+\bar{w}\chi _{n-m-1}\chi
_{m}^{\prime }\right) \right] .
\end{equation}%
\begin{figure}[th]
\includegraphics[width=0.6\textwidth]{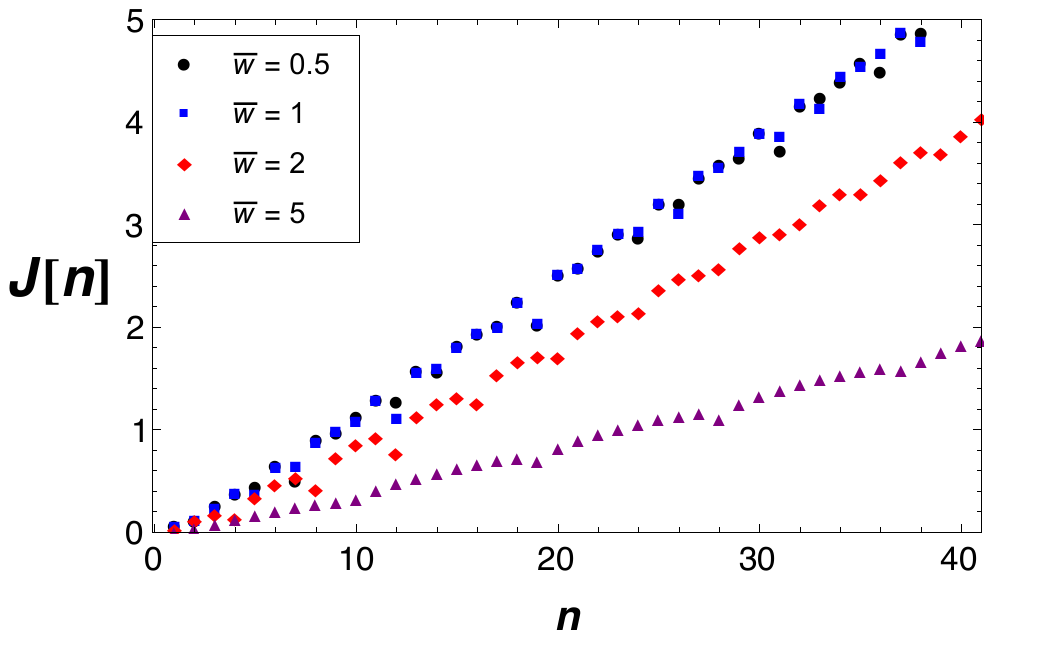}
\caption{(color online) Large order behavior of the slow-roll expansion in
conformal hydrodynamics. $J[n]=|\chi_n|^{1/n}$ as a function of $n$ with $c_{\protect\eta /R}=1/5$ and  $\bar{w}=0.5,1,2,5$.}
\label{fig:divergeslowrollCFT}
\end{figure}
The terms can be computed analytically but now the expressions are
considerably more complicated. This limits our ability to go to a very large
order in this expansion, in comparison to the constant $\tau _{R}$ case. Our
results for this series are shown in Fig.\ \ref{fig:divergeslowrollCFT} for $%
c_{R}=5c_{\eta }$. Until the order we were able to compute, the series is
found to diverge. We also checked that the same behavior holds when the
values for $c_{\eta }$ and $c_{R}$ are taken from strongly coupled $\mathcal{%
N}=4$ SYM theory \cite{Baier:2007ix}. This shows that the divergence of the
slow-roll expansion is not an exclusive feature of the set of hydrodynamic
equations obtained when the shear relaxation time is constant.

\subsection{Generalized gradient expansion}

In \cite{Denicol:2016bjh} a new type of expansion was proposed to provide a
different resummation of the famous Chapman-Enskog series for the Boltzmann
equation \cite{DeGroot:1980dk}. After just a few iterations, this new series
appeared to converge very rapidly to the exact solution for the shear stress
tensor computed using the Boltzmann equation in the relaxation time
approximation.

In this approach, the coefficients of the gradient expansion (see %
\eqref{eqdiverge}) are allowed to depend on time, i.e., we assume the
following representation for the solution 
\begin{equation}
\chi (\hat{\tau})=\frac{3a}{4}\sum_{n=0}^{\infty }\frac{c_{n}(\hat{\tau})}{%
\hat{\tau}^{n}}.  \label{newseries}
\end{equation}%
This expansion is, in principle, more general, since it allows the expansion
coefficients to have a time dependence that cannot be expanded in powers of $%
1/\hat{\tau}$. The time dependence of the generalized coefficients $c_{n}(%
\hat{\tau})$ cannot be determined a priori, but must be obtained by solving
a simple set of coupled first order linear differential equations, which can
be solved analytically. These equations are obtained by inserting %
\eqref{newseries} into \eqref{eqphi} and collecting the terms with the same
power in $1/\hat{\tau}$. This procedure rearranges the terms of the
expansion in a specific way that it naturally captures non-perturbative
exponentially small terms in Knudsen number at late times $\sim e^{-\hat{\tau%
}}$. This is mathematically justified if the series \eqref{newseries}
converges absolutely.

Another important point concerns the initial conditions. To solve the
original equation for $\chi$ one needs to specify the initial condition $%
\chi_0 \equiv \chi(\hat\tau_0)$ defined at some initial time $\hat\tau_0$.
On the other hand, since the coefficients $c_n$'s now obey first order
differential equations, one also needs to specify their initial conditions
at $\hat\tau_0$. It is natural to assume that the initial condition for the
full solution is taken care of by the zeroth order term, i.e., $c_0(\hat\tau_0)
= 4\chi_0/(3a)$, with $c_{n>0}(\hat\tau_0)=0$ - this considerably simplifies
the solutions of our hierarchy of equations order by order \cite%
{Denicol:2016bjh}. Also, it shows that \eqref{newseries} has the potential
to capture \emph{both} the late time asymptotics as well as the early time
dynamics driven by the initial condition.

In our case, the equation at zeroth order and its solution are given by 
\begin{equation}
\frac{dc_{0}}{d\hat{\tau}}+c_{0}=0\qquad \Longrightarrow \qquad c_{0}(\hat{%
\tau})=\frac{4\chi _{0}}{3a}e^{-(\hat{\tau}-\hat{\tau}_{0})}
\end{equation}%
and at first order one finds 
\begin{equation}
\frac{dc_{1}}{d\hat{\tau}}+c_{1}=1-a\,c_{0}(\hat{\tau})^{2}\qquad
\Longrightarrow \qquad c_{1}(\hat{\tau})=\left( 1-e^{-(\hat{\tau}-\hat{\tau}%
_{0})}\right) \left( 1-\frac{16\chi _{0}^{2}}{9a}e^{-(\hat{\tau}-\hat{\tau}%
_{0})}\right) .
\end{equation}%
One can see that the zeroth order solution decays exponentially in time with
a rate given by the relaxation time - this generates all the other
non-perturbative terms in Knudsen number $\sim e^{-1/K_{N}}$. The
differential equation that determines the higher order terms $(n\geq 1)$ is 
\begin{equation}
\frac{dc_{n+1}}{d\hat{\tau}}+c_{n+1}=n\,c_{n}-a\,\sum_{m=0}^{\infty
}c_{n-m}c_{m}.
\end{equation}%
This equation can be easily solved iteratively to determine the coefficients
at arbitrary order in an analytical manner. Clearly, at late times the
solutions of these equations give coefficients that are \emph{asymptotic} to
those defined by \eqref{eqdiverge}. However, we emphasize that in contrast
to the usual gradient expansion the current procedure leads to a late time
expansion that also includes exponentially small terms characteristic of
resurgent behavior.

\begin{figure}[th]
\includegraphics[width=0.6\textwidth]{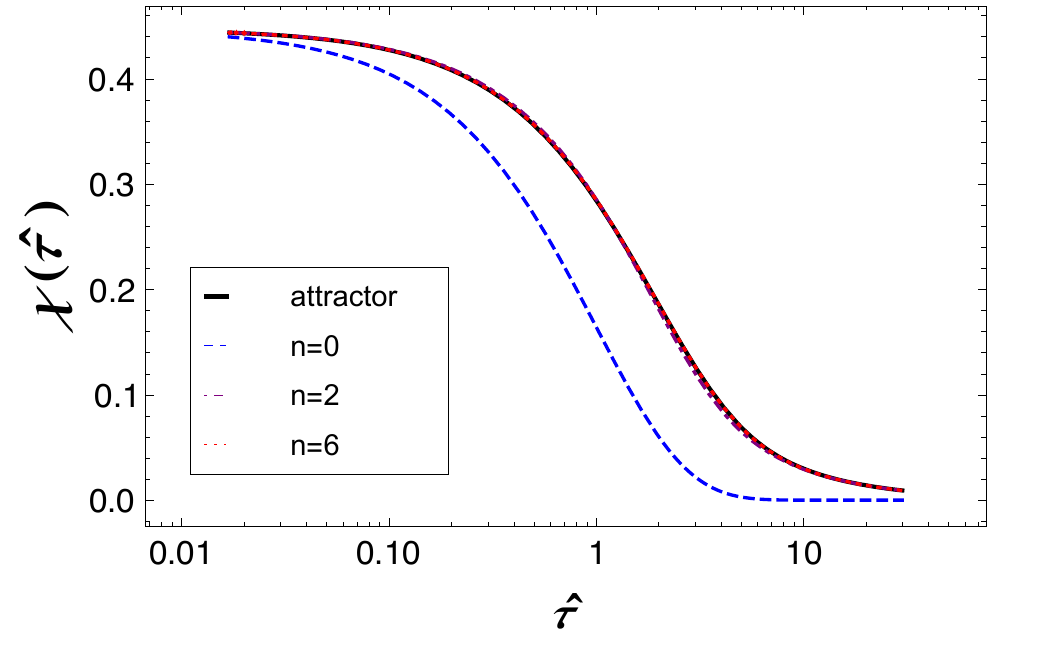}
\caption{(color online) Comparison between the analytical attractor in 
\eqref{atta1} for $a=16/45$ and the result from the generalized gradient
expansion, computed at different orders.}
\label{fig:new_expansion_comp}
\end{figure}

\begin{figure}[th]
\includegraphics[width=0.6\textwidth]{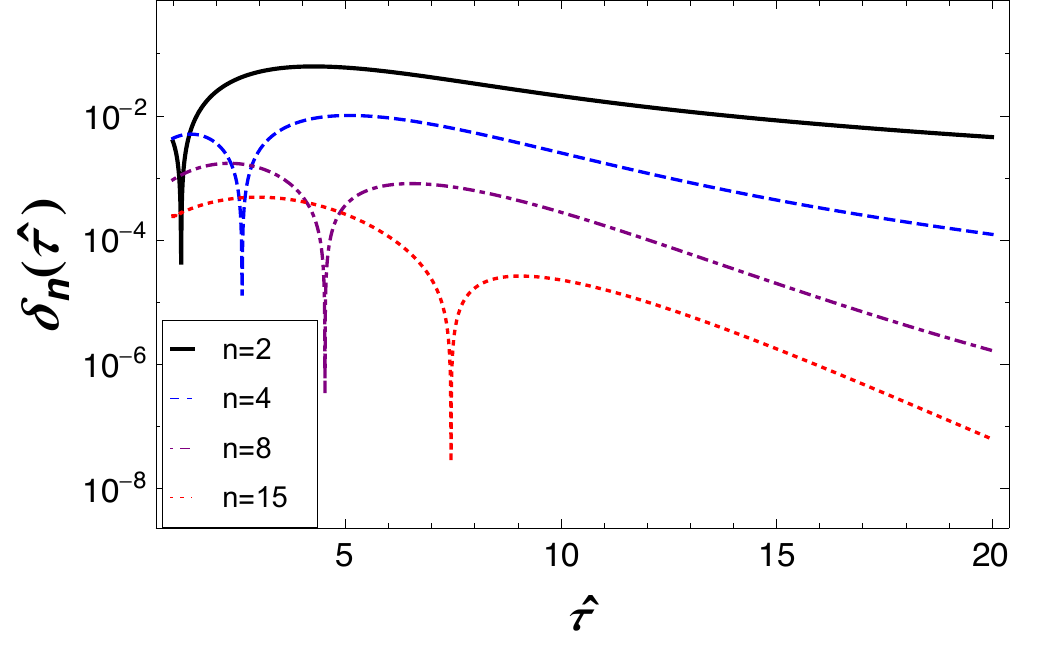}
\caption{(color online) Relative absolute difference between the analytical
attractor and the generalized gradient series for $a=16/45$ computed at
different orders.}
\label{fig:convergence_new_series}
\end{figure}

In order to investigate how this series describes the analytical attractor
in \eqref{analpi1attractor} we set $\hat{\tau}_{0}=0$ and $\chi _{0}=3\sqrt{a%
}/4$, which gives $c_{0}(\hat{\tau})=e^{-\hat{\tau}}/\sqrt{a}$ and $c_{1}(%
\hat{\tau})=(1-e^{-\hat{\tau}})^{2}$ for the first terms. We show in Fig.\ %
\ref{fig:new_expansion_comp} a comparison between the analytical attractor
and the result obtained from the new series, which approaches the analytical
solution already at second order. 
While we have not been able to verify if this new expansion converges
absolutely, in Fig.\ \ref{fig:convergence_new_series} we show that the
relative absolute difference between the analytical attractor and the new
expansion, 
\begin{equation}
\delta _{n}(\hat{\tau})=\frac{\left\vert \chi _{att}(\hat{\tau})-\frac{3a}{4}%
\sum_{m=0}^{n}\frac{c_{m}(\hat{\tau})}{\hat{\tau}^{m}}\right\vert }{\chi
_{att}(\hat{\tau})},
\end{equation}%
decreases significantly when more terms are included in the expansion (our
maximum number of terms here was 15). Even if this series is later shown to
also be divergent, one can see that it provides an excellent approximation
to the attractor already at low orders in comparison to previous approaches. 

\section{Conclusions}

\label{sec:conclusions}

In this paper we investigated the solutions of Israel-Stewart theory under
Bjorken scaling, in the absence of bulk viscous pressure contributions, at
zero chemical potential, and for a constant relaxation time. Our goal was to investigate the emergent universal behavior of these solutions at late times where all the information about the initial conditions is lost. We demonstrated that the equations of motion of Israel-Stewart theory under these conditions can be solved analytically. We determined an analytical expression for the hydrodynamic
attractor for the first time and checked if it could be reproduced, even in an approximate form, by a series expansion. In particular, we considered two expansion methods that are commonly employed in this area: the gradient expansion and the slow-roll series. 

When analyzing the gradient expansion, we confirmed that the series diverges for the values of transport coefficients that arise from the Boltzmann equation. Interestingly enough, we found that the series can converge depending on the values of $\eta/(sT\tau_R)$ and in these cases the series can even be explicitly resummed. However, we note that this was only possible for parameter choices that lead to acausal
propagation in the fluid and, consequently, are unphysical. 

More importantly, we demonstrated for the first time that the slow-roll expansion, which is widely employed to find approximate expressions for the hydrodynamic attractor, has zero
radius of convergence. This was found by showing that the terms in the series
display factorial growth, for all values of time. We note that this result also holds for a conformal fluid. Nevertheless, both series investigated have an
optimal truncation that is actually able to provide a reasonable description
of the attractor solution at late times. Therefore, these expansions can still
be used to describe the universal hydrodynamic properties of a fluid, even though they cannot be systematically improved by the inclusion of higher order terms.

Finally, we showed an example of a series that appears to converge rapidly.
This expansion was first proposed in \cite{Denicol:2016bjh}, to find
approximate solutions of the Boltzmann equation, where it also appeared to
converge. Within this approach, the coefficients of the gradient expansion
are assumed to display a non-trivial time dependence, which cannot be simply
expanded in powers of Knudsen number. Such a time dependence is obtained
directly from the equations of motion, by obtaining and solving the simple
first order differential equations satisfied by each coefficient. If this
series does in fact converge, it is currently the only option to
systematically approximate the hydrodynamic attractor of a given system. Despite the apparent success of this method, we stress that it has only been developed so far in Bjorken flow and it remains a challenge to generalize it to more general flow patterns.  

\section*{Acknowledgements}

GSD and JN thank Conselho Nacional de Desenvolvimento Cient\'{\i}fico e
Tecnol\'{o}gico (CNPq) for financial support. JN thanks Funda\c c\~ao de
Amparo \`{a} Pesquisa do Estado de S\~{a}o Paulo (FAPESP) under grant
2015/50266-2 for financial support and the Department of Physics and
Astronomy at Rutgers University for its hospitality.

\bibliography{References_attractor.bib}

\end{document}